\documentclass[english,showpacs,preprint,superscriptaddress]{revtex4-1}
\usepackage{lmodern}
\usepackage{lmodern}
\usepackage[T1]{fontenc}
\usepackage{color}
\usepackage{amsmath}
\usepackage{amssymb}
\usepackage{graphicx}
\usepackage{esint}

\makeatletter

\usepackage{amsthm}\usepackage{latexsym}\usepackage{bm}\usepackage{amsfonts}

\usepackage{babel}

\usepackage{babel}

\usepackage{hyperref}

\@ifundefined{showcaptionsetup}{}{%
 \PassOptionsToPackage{caption=false}{subfig}}
\usepackage{subfig}
\makeatother

\usepackage{babel}
\begin{document}

\title{Explicit high-order noncanonical symplectic algorithms for ideal
two-fluid systems}

\author{Jianyuan Xiao}

\affiliation{School of Nuclear Science and Technology and Department of Modern
Physics, University of Science and Technology of China, Hefei, Anhui
230026, China}

\affiliation{Key Laboratory of Geospace Environment, CAS, Hefei, Anhui 230026,
China}

\author{Hong Qin }

\email{corresponding author: hongqin@ustc.edu.cn}

\affiliation{School of Nuclear Science and Technology and Department of Modern
Physics, University of Science and Technology of China, Hefei, Anhui
230026, China}

\affiliation{Plasma Physics Laboratory, Princeton University, Princeton, NJ 08543,
U.S.}

\author{Philip J. Morrison}

\affiliation{Department of Physics and Institute for Fusion Studies, University
of Texas at Austin, Austin, TX 78741, U.S.}

\author{Jian Liu}

\affiliation{School of Nuclear Science and Technology and Department of Modern
Physics, University of Science and Technology of China, Hefei, Anhui
230026, China}

\affiliation{Key Laboratory of Geospace Environment, CAS, Hefei, Anhui 230026,
China}

\author{Zhi Yu}

\affiliation{Institute of Plasma Physics, Chinese Academy of Sciences, Hefei,
Anhui 230031, China}

\author{Ruili Zhang}

\affiliation{School of Nuclear Science and Technology and Department of Modern
Physics, University of Science and Technology of China, Hefei, Anhui
230026, China}

\affiliation{Key Laboratory of Geospace Environment, CAS, Hefei, Anhui 230026,
China}

\author{Yang He}

\affiliation{School of Nuclear Science and Technology and Department of Modern
Physics, University of Science and Technology of China, Hefei, Anhui
230026, China}

\affiliation{Key Laboratory of Geospace Environment, CAS, Hefei, Anhui 230026,
China}
\begin{abstract}
An explicit high-order noncanonical symplectic algorithm for ideal
two-fluid systems is developed. The fluid is discretized as particles
in the Lagrangian description, while the electromagnetic fields and
internal energy are treated as discrete differential form fields on
a fixed mesh. With the assistance of Whitney interpolating forms \cite{whitney1957geometric,desbrun2008discrete,xiao2015explicit},
this scheme preserves the gauge symmetry of the electromagnetic field,
and the pressure field is naturally derived from the discrete internal
energy. The whole system is solved using the Hamiltonian splitting
method discovered by \textsl{He et al.}~\cite{he2015hamiltonian},
which was been successfully adopted in constructing symplectic particle-in-cell
schemes \cite{xiao2015explicit}. Because of its structure preserving
and explicit nature, this algorithm is especially suitable for large-scale
simulations for physics problems that are multi-scale and require
long-term fidelity and accuracy. The algorithm is verified via two
tests: studies of the dispersion relation of waves in a two-fluid
plasma system and the oscillating two-stream instability. 
\end{abstract}



\maketitle
\global\long\def\EXP{\times10}
 \global\long\def\rmd{\mathrm{d}}
 \global\long\def\xs{ \mathbf{x}_{s}}
 \global\long\def\dotxs{\dot{\mathbf{x}}_{s}}
 \global\long\def\bfx{\mathbf{x}}
 \global\long\def\bfv{\mathbf{v}}
 \global\long\def\bfA{\mathbf{A}}
 \global\long\def\bfB{\mathbf{B}}
 \global\long\def\bfE{\mathbf{E}}
 \global\long\def\bfu{\mathbf{u}}
 \global\long\def\bfe{\mathbf{e}}
 \global\long\def\bfd{\mathbf{d}}
 \global\long\def\rme{\mathrm{e}}
 \global\long\def\rmi{\mathrm{i}}
 \global\long\def\rmq{\mathrm{q}}
 \global\long\def\ope{\omega_{pe}}
 \global\long\def\oce{\omega_{ce}}
 \global\long\def\FIG#1{Fig.~\ref{#1}}
 \global\long\def\EQ#1{Eq.~(\ref{#1})}
 \global\long\def\SEC#1{Sec.~\ref{#1}}
 \global\long\def\REF#1{Ref.~\cite{#1}}
 \global\long\def\CURLD{ {\mathrm{curl_{d}}}}
 \global\long\def\DIVD{ {\mathrm{div_{d}}}}
 \global\long\def\CURLDP{ {\mathrm{curl_{d}}^{T}}}
 \global\long\def\cpt{\captionsetup{justification=raggedright }}
 \global\long\def\act{\mathcal{A}}
 \global\long\def\calL{\mathcal{L}}
 \global\long\def\calJ{\mathcal{J}}

\section{Introduction}

The ideal two-fluid model, a basic non-dissipative model of plasma
physics, has been widely used to study fusion and astrophysical plasmas.
In this model, the electrons and ions are treated as ideal fluids
separately, with coupling to the electromagnetic fields through the
charge and current carried by them. Although this system is easily
generalized to any number of different charged species, the terminology
``two-fluid'' will be used here in lieu of ``multi-fluid'', as
is typically done. Because the ideal two-fluid system has noncanonical
Hamiltonian form \cite{morrison1982AIP,morrison1998hamiltonian},
as was shown in Ref.~\cite{spencer1982hamiltonian}, its dynamics
preserves geometric structure and there is no dissipation of invariants
such as the total energy and momentum in the system. Conventional
numerical algorithms for the ideal two-fluid system generally do not
preserve geometric structure and thus the truncation error can accumulate
coherently over simulation time-steps. This is a serious drawback
when solving most electron-ion systems whose behaviors are naturally
multi-scale. For example, the ion cyclotron period is thousands of
times longer than that of the electron.

Symplectic methods, discovered in the 1980s \cite{Lee82,Ruth83,Feng85,Feng86,Lee87,Veselov88,yoshida1990construction,Forest90,Channell90,Candy91,Tang93,Sanz-Serna94,Shang99,marsden2001discrete,Hairer02,Feng10},
have proven to be efficient for solving finite-dimensional canonical
Hamiltonian systems. Such methods preserve the symplectic geometric
structure (2-form) associated with the original canonical Hamiltonian
system, and the numerical error of all invariants can be globally
bounded by small values throughout simulations \cite{hairer2006geometric}.
In plasma physics, accelerator physics, and fluid dynamics, many of
the finite-dimensional Hamiltonian systems and most of the infinite-dimensional
Hamiltonian systems are noncanonical; for example, this is the case
for guiding center dynamics \cite{little1979,littlejohn1981hamiltonian,littlejohn1983},
the Euler fluid and magnetohydrodynamics (MHD) equations \cite{morrison1980noncanonical},
the Vlasov-Maxwell and Vlasov-Poisson systems \cite{morrison1980maxwell,morrison1982AIP,Weinstein81,marsden1982hamiltonian,morrison1998hamiltonian},
and drift and gyrokinetic theories \cite{pjm13,squire,pjmVG13,burby2014hamiltonian,brizard16}.
The development of geometric algorithms for these systems can be challenging.
However, recently significant advances have been achieved in the development
of structure preserving geometric algorithms for charged particle
dynamics \textcolor{black}{\cite{qin2008variational,qin2009variational,guan2010phase,squire2012gauge,qin2013boris,liu2014fate,zhang2014,zhang2015volume,ellison2015comment,he2015volume,he2015explicit,ellison2015development,liu2015neoclassical,He16-172}},
the Vlasov-Maxwell systems \textcolor{black}{\cite{Squire4748,squire2012geometric,xiao2013variational,kraus2013variational,evstatiev2013variational,Shadwick14,xiao2015variational,xiao2015explicit,crouseilles2015hamiltonian,Qin15JCP,he2015hamiltonian,qin2016canonical,Webb16}},
compressible ideal MHD \cite{zhou2014variational,zhou2015formation},
and incompressible fluids \cite{pavlov2011structure,gawlik2011geometric}.
All of these methods have demonstrated unparalleled long-term numerical
accuracy and fidelity compared with conventional methods. As a side
note, we point out that for infinite-dimensional Hamiltonian systems,
an alternative viewpoint is to treat them as multi-symplectic systems
\cite{Bridges97,marsden1998multisymplectic}, and corresponding multi-symplectic
algorithms \cite{Reich00,Sun00,bridges2001multi,Wang01,Hong02,Chen02,Guo03}
have also been developed.

In the present work, an explicit, high-order, noncanonical symplectic
algorithm for integrating the compressible ideal two-fluid system
is developed. We discretize the fluid as particles in the Lagrangian
description, which naturally guarantees conservation of the density.
The electromagnetic fields and internal energy are discretized over
a cubic mesh by using the theory of discrete exterior calculus (DEC)
\cite{hirani2003discrete}. High-order Whitney interpolating forms
\cite{xiao2015explicit} are used to ensure the gauge symmetry of
Maxwell's equations. The discrete Poisson bracket for the ideal two-fluid
system is obtained by the similar technique that is used in obtaining
the discrete Vlasov-Maxwell bracket \cite{xiao2015explicit}, and
the final numerical scheme is constructed by the powerful Hamiltonian
split method \cite{he2015hamiltonian,he2015explicit,xiao2015explicit}.
We note that for the existing structure preserving method for the
compressible fluid \cite{zhou2014variational}, all fields are discretized
over a moving mesh, which does not apply to cases where the mesh deforms
significantly during the evolution, such as in a rotating fluid. This
difficulty is overcome by using a fixed mesh rather than a moving
one for discretizing the electromagnetic and internal energy fields
in our method. The conservation of symplectic structure guarantees
that the numerical errors of all invariants such as the total energy
and momentum are bounded within a small value during the simulations
\cite{hairer2006geometric}. Therefore, this method is most suitable
for solving long-term multi-scale problems.

The paper is organized as follows. In Sec.~\ref{SecTheory} the Hamiltonian
theory of the ideal two-fluid system is reviewed and the geometric
structure preserving method is developed. Two numerical examples,
the dispersion relation of waves in an ideal two-fluid system and
the oscillating two-stream instability, are given in Sec.~\ref{SecNumer}.
Finally, in Sec.~\ref{SecDis} we conclude.

\section{Structure preserving discretization for ideal two-fluid systems}

\label{SecTheory}

The starting point of our development is the Lagrangian of the ideal
two-fluid system, written in terms of Lagrangian variables, which
is quite similar to the Lagrangian for the Vlasov-Maxwell system except
for the addition of internal energy terms (see e.g.\ \cite{charidakos2014action}).
This Lagrangian is given as follows: 
\begin{eqnarray}
\notag\calL & = & \sum_{s}\int\rmd\bfx_{0}\left(\frac{1}{2}n_{s0}(\mathbf{x}_{0})m_{s}|\dot{\bfx}_{s}|^{2}+q_{s}n_{s0}(\mathbf{x}_{0})\dot{\bfx}_{s}\cdot\bfA\left(\bfx_{s}\right)-U_{ms}\left(\frac{n_{s0}(\mathbf{x}_{0})m_{s}}{\calJ\left(\bfx_{s}\right)}\right)\right)\\
 &  & \hspace{2cm}+\ \frac{1}{2}\int\rmd^{3}\bfx\left(\left|\dot{\bfA}\left(\bfx\right)\right|^{2}-\left|\nabla\times\bfA\left(\bfx\right)\right|^{2}\right)\,,
\end{eqnarray}
where $m_{s}$, $q_{s}$, and $n_{s0}$ are the mass, charge, and
initial number density distribution of species $s$, respectively,
$\mathbf{x}_{s}$ and $\dot{\mathbf{x}}_{s}$ are current position
and velocity of fluid elements for species $s$ labeled by $\bfx_{0}$,
which we take to be the initial value of $\mathbf{x}_{s}$ in the
configuration space, $\calJ\left(\bfx_{s}\right)$ is the Jacobian
of the coordinate transformation from the initial value $\bfx_{0}$
to $\bfx_{s}$, $U_{ms}$ is the internal energy per unit mass for
species $s$, and $\bfA$ is the electromagnetic vector potential.
In the arguments of the fields $\mathbf{x}_{s}$ and $\bfA$ we suppress
the time variable. In this Lagrangian, we have ignored the entropy
term in the internal energy, assuming barotropic fluids, and adopted
the temporal gauge with $\phi=0$. The permittivity and permeability
are set to unity for simplicity.

Evolution equations are obtained upon variation of the action $S$
as in Hamilton's principle 
\begin{equation}
\delta S[\mathbf{x}_{s},\bfA]=\delta\int\!\rmd t\,\calL=0
\end{equation}
giving rise the equations of motion via 
\begin{equation}
\frac{\delta S}{\delta\bfx_{s}}=0\qquad\mathrm{and}\qquad\frac{\delta S}{\delta\bfA}=0\,,
\end{equation}
which yield 
\begin{eqnarray}
m_{s}\ddot{\bfx}_{s}(\mathbf{x}_{0}) & = & q_{s}\big(\dot{\bfx}_{s}\times\bfB\left(\bfx_{s}\right)+\bfE\left(\bfx_{s}\right)\big)-\frac{1}{n_{s0}(\mathbf{x}_{0})}\frac{\partial U_{ms}\left(\frac{n_{s0}(\mathbf{x}_{0})m_{s}}{\calJ\left(\bfx_{s}\right)}\right)}{\partial\bfx_{s}}~,\\
\dot{\bfE}\left(\bfx\right) & = & \nabla\times\bfB\left(\bfx\right)-\sum_{s}\int\!\rmd\bfx_{0}\,q_{s}n_{s0}(\mathbf{x}_{0})\dot{\bfx}_{s}\delta\left(\bfx-\bfx_{s}\right)~,
\end{eqnarray}
where $\bfE=-\dot{\bfA}$ and $\bfB=\nabla\times\bfA$ are the electromagnetic
fields. These equations are exactly the ideal two-fluid equations
in the Lagrangian variable description.

Now we discretize the Lagrangian using a method very similar to that
for the discretization of the Vlasov-Maxwell system in Ref.~\cite{xiao2015explicit}.
The electromagnetic fields and internal energy are sampled over a
cubic mesh, while the fluid is discretized into finite-sized smooth
particle \cite{monaghan1992smoothed,xiao2013variational,xiao2015explicit}
moving between mesh grids. Modeling fluids using a set of Lagrangian
particles is also the key idea of the smoothed-particle-hydrodynamics
(SPH) method \cite{monaghan1992smoothed,bonet1999variational,price2004smoothed}.
However, the difference is that our internal energy fields are calculated
on fixed mesh grids. Therefore, the method developed in the present
study more closely resembles the structure-preserving symplectic particle-in-cell
method of Ref.~\cite{xiao2015explicit} The resulting discrete Lagrangian
is 
\begin{eqnarray}
\calL_{d} & = & \sum_{s,p}\left(\frac{1}{2}m_{s}n_{0,sp}|\dot{\bfx}_{sp}|^{2}+q_{s}n_{0,sp}\dot{\mathbf{x}}_{sp}\cdot\sum_{J}\bfA_{J}W_{\sigma_{1J}}\left(\bfx_{sp}\right)\right)-\sum_{s,I}U_{s}\left(\rho_{sI}\right)\nonumber \\
 &  & \hspace{2cm}+\ \frac{1}{2}\sum_{J}\left(\left|\dot{\bfA}_{J}\right|^{2}-\left|\CURLD\bfA_{J}\right|^{2}\right)\,,\label{Ld}
\end{eqnarray}
where 
\begin{eqnarray}
\rho_{sI} & = & \sum_{p}m_{s}n_{0,sp}W_{\sigma_{0I}}\left(\bfx_{sp}\right)\,.
\end{eqnarray}
Here, the subscript $sp$ denotes the $p$-th particle of species
$s$, $W_{\sigma_{0I}}$ and $W_{\sigma_{1J}}$ are Whitney interpolating
maps for discrete 0-forms and 1-forms \cite{hirani2003discrete,whitney1957geometric,desbrun2008discrete,xiao2015explicit},
$U_{s}$ is discrete internal energy per unit volume for species $s$,
$\CURLD$ is the discrete curl operator that is defined in Eq.~(\ref{eq:DEFCURLD}),
$I,\thinspace J,\thinspace K$ are indices for the discrete 0-form,
1-form, 2-form, respectively. To simplify the notation, the grid size
$\Delta x$ has been set to unity. The Whitney maps are defined as
follows: 
\begin{eqnarray*}
\sum_{i,j,k}W_{\sigma_{0,i,j,k}}\left(\bfx\right)\phi_{i,j,k} & \equiv & \sum_{i,j,k}\phi_{i,j,k}W_{1}\left(x\right)W_{1}\left(y\right)W_{1}\left(z\right),\\
\sum_{i,j,k}W_{\sigma_{1,i,j,k}}\left(\bfx\right)\bfA_{i,j,k} & \equiv & \sum_{i,j,k}\left[\begin{array}{c}
{A_{x}}_{i,j,k}W_{1}^{(2)}(x-i)W_{1}(y-j)W_{1}(z-k)\\
{A_{y}}_{i,j,k}W_{1}(x-i)W_{1}^{(2)}(y-j)W_{1}(z-k)\\
{A_{z}}_{i,j,k}W_{1}(x-i)W_{1}(y-j)W_{1}^{(2)}(z-k)~
\end{array}\right]^{T},\\
\sum_{i,j,k}W_{\sigma_{2,i,j,k}}\left(\bfx\right)\bfB_{i,j,k} & \equiv & \sum_{i,j,k}\left[\begin{array}{c}
{B_{x}}_{i,j,k}W_{1}(x-i)W_{1}^{(2)}(y-j)W_{1}^{(2)}(z-k)\\
{B_{y}}_{i,j,k}W_{1}^{(2)}(x-i)W_{1}(y-j)W_{1}^{(2)}(z-k)\\
{B_{z}}_{i,j,k}W_{1}^{(2)}(x-i)W_{1}^{(2)}(y-j)W_{1}(z-k)
\end{array}\right]^{T},\\
\sum_{i,j,k}W_{\sigma_{3,i,j,k}}\left(\bfx\right)\rho_{i,j,k} & \equiv & \sum_{i,j,k}\rho_{i,j,k}W_{1}^{(2)}(x-i)W_{1}^{(2)}(y-j)W_{1}^{(2)}(z-k),\\
W_{1}^{(2)}\left(x\right) & = & -\left\{ \begin{array}{lc}
W_{1}'\left(x\right)+W_{1}'\left(x+1\right)+W_{1}'\left(x+2\right)~, & -1\leq x<2~,\\
0~, & \mathrm{otherwise}~.
\end{array}\right.
\end{eqnarray*}
where the one-dimensional interpolation function $W_{1}$ is chosen
in this paper to be 
\begin{equation}
W_{1}(x)=\begin{cases}
0, & x\leq-2,\\
-\frac{{x^{6}}}{{48}}-\frac{{x^{5}}}{{8}}-\frac{{5\,x^{4}}}{{16}}-\frac{{5\,x^{3}}}{{12}}+x+1, & -2<x\leq-1,\\
\frac{{x^{6}}}{{48}}-\frac{{x^{5}}}{{8}}-\frac{{5\,x^{4}}}{{16}}-\frac{{5\,x^{3}}}{{12}}-\frac{{5\,x^{2}}}{{8}}+\frac{{7}}{{12}}, & -1<x\leq0,\\
\frac{{x^{6}}}{{48}}+\frac{{x^{5}}}{{8}}-\frac{{5\,x^{4}}}{{16}}+\frac{{5\,x^{3}}}{{12}}-\frac{{5\,x^{2}}}{{8}}+\frac{{7}}{{12}}, & 0<x\leq1,\\
-\frac{{x^{6}}}{{48}}+\frac{{x^{5}}}{{8}}-\frac{{5\,x^{4}}}{{16}}+\frac{{5\,x^{3}}}{{12}}-x+1, & 1<x\leq2,\\
0, & 2<x\,.
\end{cases}
\end{equation}

The equations of motion arising from the action with Lagrangian $\calL_{d}$
of \eqref{Ld} are the following: 
\begin{eqnarray}
\notag m_{s}n_{0,sp}\ddot{\bfx}_{sp} & = & q_{s}n_{0,sp}\left(\dot{\bfx}_{sp}\times\left(\nabla\times\sum_{J}\bfA_{J}W_{\sigma_{1J}}\left(\bfx_{sp}\right)\right)-\dot{\bfA}_{J}W_{\sigma_{1J}}\left(\bfx_{sp}\right)\right)\\
 &  & \hspace{1cm}-\ \sum_{I}U_{s}'\left(\rho_{sI}\right)m_{s}n_{0,sp}\nabla W_{\sigma_{0I}}\left(\bfx_{sp}\right)~,\label{eq:mnx}\\
\ddot{\bfA}_{J} & = & -\ \CURLDP\CURLD\bfA_{J}+\sum_{s,p}q_{s}n_{0,sp}\bfx_{sp}W_{\sigma_{1J}}\left(\bfx_{sp}\right)\,.\label{eq:Aj}
\end{eqnarray}
Next we introduce two discrete fields $\bfE_{J}=-\dot{\bfA}_{J}$
and $\bfB_{K}=\sum_{J}\CURLD_{KJ}\bfA_{J}$, which are discrete electromagnetic
fields. We will make use of the following properties of the interpolating
forms \cite{xiao2015explicit,hirani2003discrete,whitney1957geometric},
\begin{eqnarray}
\nabla\sum_{I}W_{\sigma_{0I}}\left(\bfx\right) & \phi_{I}= & \sum_{I,J}W_{\sigma_{1J}}\left(\bfx\right){\nabla_{\mathrm{d}}}_{JI}\phi_{I}~,\label{EqnD0to1FORMAPP}\\
\nabla\times\sum_{J}W_{\sigma_{1J}}\left(\bfx\right)\bfA_{J} & = & \sum_{J,K}W_{\sigma_{2K}}\left(\bfx\right){\CURLD}_{KJ}\bfA_{J}~,\label{EqnD1to2FORMAPP}\\
\nabla\cdot\sum_{K}W_{\sigma_{2K}}\left(\bfx\right)\bfB_{K} & = & \sum_{K,L}W_{\sigma_{3L}}\left(\bfx\right){\DIVD}_{LK}\bfB_{K}~,\label{EqnD2to3FORMAPP}\\
\left({\nabla_{\mathrm{d}}}\phi\right)_{i,j,k} & = & [\phi_{i+1,j,k}-\phi_{i,j,k},\phi_{i,j+1,k}-\phi_{i,j,k},\phi_{i,j,k+1}-\phi_{i,j,k}]~.\label{EqnDEFGRADD}\\
\left(\CURLD\bfA\right)_{i,j,k} & = & \left[\begin{array}{c}
\left({A_{z}}_{i,j+1,k}-{A_{z}}_{i,j,k}\right)-\left({A_{y}}_{i,j,k+1}-{A_{y}}_{i,j,k}\right)\\
\left({A_{x}}_{i,j,k+1}-{A_{x}}_{i,j,k}\right)-\left({A_{z}}_{i+1,j,k}-{A_{z}}_{i,j,k}\right)\\
\left({A_{y}}_{i+1,j,k}-{A_{y}}_{i,j,k}\right)-\left({A_{x}}_{i,j+1,k}-{A_{x}}_{i,j,k}\right)
\end{array}\right]^{T}~,\label{eq:DEFCURLD}\\
\left({\DIVD}\bfB\right)_{i,j,k} & = & \left({B_{x}}_{i+1,j,k}-{B_{x}}_{i,j,k}\right)+\left({B_{y}}_{i,j+1,k}-{B_{y}}_{i,j,k}\right)\notag\\
 &  & \hspace{2cm}+\ \left({B_{z}}_{i,j,k+1}-{B_{z}}_{i,j,k}\right)~.\label{EqnDEFDIVD}
\end{eqnarray}
which hold for any $\phi_{I}$, $\bfA_{J}$, and $\bfB_{K}$. With
these identities, Eqs.~\eqref{eq:mnx} and \eqref{eq:Aj} can be
expressed as 
\begin{eqnarray}
\notag m_{s}n_{0,sp}\ddot{\bfx}_{sp} & = & q_{s}n_{0,sp}\left(\dot{\bfx}_{sp}\times\sum_{K}\bfB_{K}W_{\sigma_{2K}}\left(\bfx_{sp}\right)+\bfE_{J}W_{\sigma_{1J}}\left(\bfx_{sp}\right)\right)\\
 &  & \hspace{2cm}-\sum_{I}U_{i}'\left(\rho_{sI}\right)m_{s}n_{0,sp}\nabla W_{\sigma_{0I}}\left(\bfx_{sp}\right)~,\label{EqnMiORI}\\
\dot{\bfE}_{J} & = & \sum_{K}\CURLDP_{JK}\bfB_{K}-\sum_{s,p}q_{s}n_{0,sp}\dot{\mathbf{x}}_{sp}W_{\sigma_{1J}}\left(\bfx_{sp}\right)~,\label{EqnEORI}\\
\dot{\bfB}_{K} & = & -\sum_{J}\CURLD_{KJ}\bfE_{J}~.\label{EqnBORI}
\end{eqnarray}
The continuity equations for the densities are automatically satisfied,
as can be shown by directly calculating the time derivative of $\rho_{sI}$,

\begin{eqnarray}
\dot{\rho}_{sI} & = & \sum_{p}m_{s}n_{0,sp}\dot{\bfx}_{sp}\cdot\nabla W_{\sigma_{0I}}\left(\bfx_{sp}\right)\\
 & = & \sum_{p}m_{s}n_{0,sp}\dot{\bfx}_{sp}\cdot\sum_{J}W_{\sigma_{1J}}\left(\bfx_{sp}\right){\nabla_{\mathrm{d}}}_{JI}\\
 & = & \sum_{J}{\nabla_{\mathrm{d}}}_{JI}\sum_{p}m_{s}n_{0,sp}\dot{\bfx}_{sp}\cdot W_{\sigma_{1J}}\left(\bfx_{sp}\right)\\
 & = & \sum_{J}{\nabla_{\mathrm{d}}}_{JI}\mathbf{M}_{J}~,\label{eq:DMJ}
\end{eqnarray}
where $\mathbf{M}_{J}$ can be viewed as the discrete momentum density
over mesh grids, and $\sum_{J}{\nabla_{\mathrm{d}}}_{JI}$ is a discrete
version of $-\nabla\cdot$. So, Eq.\,\eqref{eq:DMJ} is essentially
a kind of discrete continuity equation.

To construct the geometric structure preserving algorithm, the Hamiltonian
theory for the discretized system is considered. Note that the only
difference between the two-fluid Lagrangian and the Vlasov-Maxwell
Lagrangian is the internal energy term, which can be written as a
function of $\bfx_{sp}$. Thus, the discrete Poisson structure of
the ideal two-fluid system can be chosen to be the same as that for
the Vlasov-Maxwell system \cite{xiao2015explicit}, which is 
\begin{eqnarray}
\notag\left\{ F,G\right\}  & = & \sum_{J}\left(\frac{\partial F}{\partial\bfE_{J}}\cdot\sum_{K}\frac{\partial G}{\partial\bfB_{K}}\CURLD_{KJ}-\sum_{K}\frac{\partial F}{\partial\bfB_{K}}\CURLD_{KJ}\cdot\frac{\partial G}{\partial\bfE_{J}}\right)\\
\notag &  & +\ \sum_{s,p}\frac{1}{m_{s}n_{0,sp}}\left(\frac{\partial F}{\partial\bfx_{sp}}\cdot\frac{\partial G}{\partial\dot{\bfx}_{sp}}-\frac{\partial F}{\partial\dot{\bfx}_{sp}}\cdot\frac{\partial G}{\partial\bfx_{sp}}\right)\\
\notag &  & +\ \sum_{s,p}\frac{q_{s}}{m_{s}}\left(\frac{\partial F}{\partial\dot{\bfx}_{sp}}\cdot\sum_{J}W_{\sigma_{1J}}\left(\bfx_{sp}\right)\frac{\partial G}{\partial\bfE_{J}}-\frac{\partial G}{\partial\dot{\bfx}_{sp}}\cdot\sum_{J}W_{\sigma_{1J}}\left(\bfx_{sp}\right)\frac{\partial F}{\partial\bfE_{J}}\right)\\
\notag &  & -\ \sum_{s,p}\frac{q_{s}}{m_{s}^{2}n_{0,sp}}\frac{\partial F}{\partial\dot{\bfx}_{sp}}\cdot\left[\sum_{K}W_{\sigma_{2K}}\left(\bfx_{sp}\right)\bfB_{K}\right]\times\frac{\partial G}{\partial\dot{\bfx}_{sp}}~.
\end{eqnarray}
And the two-fluid Hamiltonian is 
\begin{eqnarray}
H & = & \frac{1}{2}\left(\sum_{J}\bfE_{J}^{2}+\sum_{K}\bfB_{K}^{2}+\sum_{s,p}m_{s}n_{0,sp}|\dot{\bfx}_{sp}|^{2}\right)+\sum_{sI}U_{s}\left(\rho_{sI}\right)\thinspace.\label{eq:EqnHalmitEB}
\end{eqnarray}
It is straightforward to check that the following Hamiltonian equations
are identical to Eqs.~(\ref{EqnMiORI}-\ref{EqnBORI}), 
\begin{eqnarray}
\dot{\bfx}_{sp} & = & \left\{ {\bfx}_{sp},H\right\} ~,\\
\ddot{\bfx}_{sp} & = & \left\{ \dot{\bfx}_{sp},H\right\} ~,\\
\dot{\bfE}_{J} & = & \left\{ \bfE_{J},H\right\} ~,\\
\dot{\bfB}_{K} & = & \left\{ \bfB_{K},H\right\} ~.
\end{eqnarray}

\global\long\def\bracketmJ{\left( \frac{m_{s}n_{s0}}{\calJ} \right)}

Now the discrete algorithm can be developed. Using a Hamiltonian splitting
technique similar to that in Ref.~\cite{he2015hamiltonian,xiao2015explicit},
$H$ can be split into 6 parts 
\begin{eqnarray}
H=H_{E}+H_{B}+H_{x}+H_{y}+H_{z}+H_{U}~,
\end{eqnarray}
where 
\begin{eqnarray}
H_{E} & = & \frac{1}{2}\sum_{J}\bfE_{J}^{2}~,\\
H_{B} & = & \frac{1}{2}\sum_{K}\bfB_{K}^{2}~,\\
H_{r} & = & \frac{1}{2}\sum_{s,p}m_{s}n_{0,sp}\dot{r}_{sp}^{2}\thinspace,\quad\textrm{for }r\in\{x,y,z\}\,,\\
H_{U} & = & \sum_{s,I}U_{s}\left(\rho_{sI}\right)~.
\end{eqnarray}
It turns out that the exact solutions for all sub-systems can be found
and computed explicitly. The exact solutions for $H_{E}$, $H_{B}$,
$H_{x}$, $H_{y}$ and $H_{z}$ have been derived in Ref.~\cite{xiao2015explicit}.
They are 
\begin{eqnarray}
\Theta_{E}: &  & \left\{ \begin{array}{ccl}
\bfE_{J}\left(t+\Delta t\right) & = & \bfE_{J}\left(t\right)~,\\
\bfB_{K}\left(t+\Delta t\right) & = & \bfB_{K}\left(t\right)-\Delta t\sum_{J}\CURLD_{KJ}\bfE_{J}(t)~,\\
\bfx_{sp}\left(t+\Delta t\right) & = & \bfx_{sp}\left(t\right)~,\\
\dot{\bfx}_{sp}\left(t+\Delta t\right) & = & \dot{\bfx}_{sp}\left(t\right)+\frac{q_{s}}{m_{s}}\Delta t\sum_{J}W_{\sigma_{1J}}\left(\bfx_{sp}(t)\right)\bfE_{J}(t)~,
\end{array}\right.\\
\Theta_{B}: &  & \left\{ \begin{array}{ccl}
\bfE_{J}\left(t+\Delta t\right) & = & \bfE_{J}\left(t\right)+\Delta t\sum_{K}\CURLD_{KJ}\bfB_{K}(t)~,\\
\bfB_{K}\left(t+\Delta t\right) & = & \bfB_{K}\left(t\right)~,\\
\bfx_{sp}\left(t+\Delta t\right) & = & \bfx_{sp}\left(t\right)~,\\
\dot{\bfx}_{sp}\left(t+\Delta t\right) & = & \dot{\bfx}_{sp}\left(t\right)~,
\end{array}\right.\\
\Theta_{x}: &  & \left\{ \begin{array}{ccl}
\bfE_{J}\left(t+\Delta t\right) & = & \bfE_{J}\left(t\right)-\int_{0}^{\Delta t}dt'\sum_{s,p}q_{s}n_{0,sp}\dot{x}_{sp}(t)\bfe_{x}W_{\sigma_{1J}}\left(\bfx_{sp}\left(t\right)+\dot{x}_{sp}(t)t'\bfe_{x}\right)-\\
\bfB_{K}\left(t+\Delta t\right) & = & \bfB_{K}\left(t\right)~,\\
\bfx_{sp}\left(t+\Delta t\right) & = & \bfx_{sp}\left(t\right)+\Delta t\dot{x}_{sp}(t)\bfe_{x}~,\\
\dot{\bfx}_{sp}\left(t+\Delta t\right) & = & \dot{\bfx}_{sp}\left(t\right)+\frac{q_{s}}{m_{s}}\dot{x}_{sp}(t)\bfe_{x}\times\int_{0}^{\Delta t}dt'\sum_{K}W_{\sigma_{2K}}\left(\bfx_{sp}\left(t\right)+\dot{x}_{sp}(t)t'\bfe_{x}\right)\bfB_{K}(t)~.
\end{array}\right.
\end{eqnarray}
The solutions $\Theta_{y}$ and $\Theta_{z}$ are similar to $\Theta_{x}$.
For $H_{U}$, the exact evolution equations are 
\begin{eqnarray}
\dot{\bfE}_{J} & = & \left\{ \bfE_{J},H_{U}\right\} =0~,\\
\dot{\bfB}_{K} & = & \left\{ \bfB_{K},H_{U}\right\} =0~,\\
\dot{\bfx}_{sp} & = & \left\{ \bfx_{sp},H_{U}\right\} =0~,\\
\ddot{\bfx}_{sp} & = & \left\{ \dot{\bfx}_{sp},H_{U}\right\} =-\sum_{I}U_{s}'\left(\rho_{sI}\right)\nabla W_{\sigma_{0I}}\left(\bfx_{sI}\right)\,.
\end{eqnarray}
Using the property \EQ{EqnD0to1FORMAPP} of the Whitney interpolating
forms, the exact solution can be written as 
\begin{eqnarray}
\Theta_{U}:\left\{ \begin{array}{ccl}
\bfE_{J}\left(t+\Delta t\right) & = & \bfE_{J}\left(t\right)~,\\
\bfB_{K}\left(t+\Delta t\right) & = & \bfB_{K}\left(t\right)~,\\
\bfx_{sp}\left(t+\Delta t\right) & = & \bfx_{sp}\left(t\right)~,\\
\dot{\bfx}_{sp}\left(t+\Delta t\right) & = & \dot{\bfx}_{sp}\left(t\right)-\Delta t\sum_{I,J}{\nabla_{\mathrm{d}}}_{JI}U_{s}'\left(\rho_{sI}\right)W_{\sigma_{1J}}\left(\bfx_{sp}\right)\,.
\end{array}\right.
\end{eqnarray}
Here, $\Theta_{U}$ can be interpreted as a discrete version of the
continuous Newton's second law, 
\begin{eqnarray}
\ddot{\bfx}_{s} & = & -\nabla_{\mathbf{x}_{s}}U_{s}'\left(\frac{m_{s}n_{s0}}{\calJ_{s}}\right)\,.
\end{eqnarray}
At first look, it seems different from Newton's law in the Lagrangian
form derived in Ref.~\cite{morrison1998hamiltonian}, i.e., 
\begin{eqnarray}
\ddot{\bfx} & = & -\calJ\nabla\left(\frac{\rho_{0}}{\calJ^{2}}U_{m}'\left(\frac{\rho_{0}}{\calJ}\right)\right)~.\label{EqnMorrisonPressure}
\end{eqnarray}
This is because the $U_{m}$ in \EQ{EqnMorrisonPressure} is defined
as the internal energy per mass. Upon letting $\rho_{0}=m_{s}n_{s0}$,
the relation between $U_{ms}$ and $U_{s}$ is 
\begin{eqnarray}
U_{s}\left(\frac{m_{s}n_{s0}}{\calJ}\right)=\frac{m_{s}n_{s0}}{\calJ}U_{ms}\left(\frac{m_{s}n_{s0}}{\calJ}\right)~.
\end{eqnarray}
Consequently, 
\begin{eqnarray}
\notag-\nabla U_{s}'\bracketmJ & = & -\nabla\left(U_{ms}\bracketmJ+\frac{m_{s}n_{s0}}{\calJ}U_{ms}'\bracketmJ\right)\\
\notag & = & \nabla\calJ\frac{m_{s}n_{s0}}{\calJ^{2}}U_{ms}'\bracketmJ-\nabla\left(\frac{m_{s}n_{s0}}{\calJ}U_{ms}'\bracketmJ\right)\\
\notag & = & -\calJ\nabla\left(\frac{m_{s}n_{s0}}{\calJ^{2}}U_{ms}'\bracketmJ\right)~,
\end{eqnarray}
which is identical to the pressure term in the right hand side of
\EQ{EqnMorrisonPressure}. Therefore, the pressure for the species
$s$ can be defined to be \cite{morrison1998hamiltonian} 
\begin{eqnarray}
\notag P_{s} & = & \frac{m_{s}^{2}n_{s0}^{2}}{\calJ^{2}}U_{ms}'\bracketmJ\\
\notag & = & \frac{m_{s}n_{s0}}{\calJ}U_{s}'\bracketmJ-U_{s}\bracketmJ\\
\notag & = & \rho_{s}U_{s}'\left(\rho_{s}\right)-U_{s}\left(\rho_{s}\right)~.
\end{eqnarray}

The final geometric structure-preserving scheme can be constructed
from these exact solutions. For example, a first-order scheme can
be chosen as 
\begin{eqnarray}
\Theta_{1}\left(\Delta t\right)=\Theta_{E}\left(\Delta t\right)\Theta_{B}\left(\Delta t\right)\Theta_{x}\left(\Delta t\right)\Theta_{y}\left(\Delta t\right)\Theta_{z}\left(\Delta t\right)\Theta_{U}\left(\Delta t\right)~,
\end{eqnarray}
and a second-order scheme can be constructed as 
\begin{eqnarray}
\Theta_{2}\left(\Delta t\right) & = & \Theta_{x}\left(\Delta t/2\right)\Theta_{y}\left(\Delta t/2\right)\Theta_{z}\left(\Delta t/2\right)\Theta_{B}\left(\Delta t/2\right)\Theta_{U}\left(\Delta/2\right)\Theta_{E}\left(\Delta t\right)\nonumber \\
 &  & \Theta_{U}\left(\Delta t/2\right)\Theta_{B}\left(\Delta t/2\right)\Theta_{z}\left(\Delta t/2\right)\Theta_{y}\left(\Delta t/2\right)\Theta_{x}\left(\Delta t/2\right)~.
\end{eqnarray}
The $2\left(l+1\right)$th-order scheme can be derived from the $2l$th-order
scheme by using 
\begin{eqnarray}
\Theta_{2(l+1)}(\Delta t) & = & \Theta_{2l}(\alpha_{l}\Delta t)\Theta_{2l}(\beta_{l}\Delta t)\Theta_{2l}(\alpha_{l}\Delta t)~,\\
\alpha_{l} & = & 1/(2-2^{1/(2l+1)})~,\\
\beta_{l} & = & 1-2\alpha_{l}~.
\end{eqnarray}

\section{Numerical examples}

\label{SecNumer}

To verify the practicability of our explicit high-order noncanonical
symplectic algorithm for ideal two-fluid systems, we apply it to two
physics problems. In the first problem we examine the dispersion relation
of an electron-deuterium plasma, while the second concerns the oscillation
two-stream instability.

For the electron-deuterium plasma, parameters of the unperturbed uniform
plasma are chosen as follows, 
\begin{eqnarray}
n_{i_{0}}=n_{e_{0}} & = & 4.0\EXP^{{19}}\mathrm{m}^{-3}~,\\
\rho_{e_{0}} & = & n_{e_{0}}m_{e}~,\\
\rho_{i_{0}} & = & n_{i_{0}}m_{i}~,\\
m_{i}=3671m_{e} & = & 3.344\EXP^{-27}\mathrm{kg}~,\\
q_{i}=-q_{e} & = & 1.602\EXP^{-19}\mathrm{C}~,\\
U_{e}\left(\rho_{e}\right) & = & U_{e_{0}}\frac{3}{2}\left(\frac{\rho_{e}}{\rho_{e_{0}}}\right)^{5/3}~,\\
U_{i}\left(\rho_{i}\right) & = & U_{i_{0}}\frac{3}{2}\left(\frac{\rho_{i}}{\rho_{i_{0}}}\right)^{5/3}~,\\
\bfB_{0} & = & \left(3.17\bfe_{z}+1.13\bfe_{y}\right)\mathrm{T}~,
\end{eqnarray}
where $U_{i_{0}}=U_{e_{0}}=m_{e}n_{e}v_{Te}^{2}/2,v_{Te}=0.04472\mathrm{c}$,
$\mathrm{c}$ is the speed of light in the vacuum, $\bfB_{0}$ is
the constant external magnetic field. This plasma supports both electron
waves and ion waves, and their frequencies are very different since
the deuterium ion is much heavier than the electron. The simulation
is carried out in a $1\times1\times1536$ mesh, and the periodical
boundary condition is adopted in all $x,y,z$ directions. The grid
size is choosen to be $\Delta x=2\EXP^{-4}\mathrm{m}$, the time step
is set to $\Delta t=\Delta x/\left(2\mathrm{c}\right)$. The simulation
is initialized with stationary fluid particles being equally spaced
with 4 particles per grid cell.

\begin{figure}
\subfloat[High frequency region.]{\includegraphics[width=0.49\textwidth]{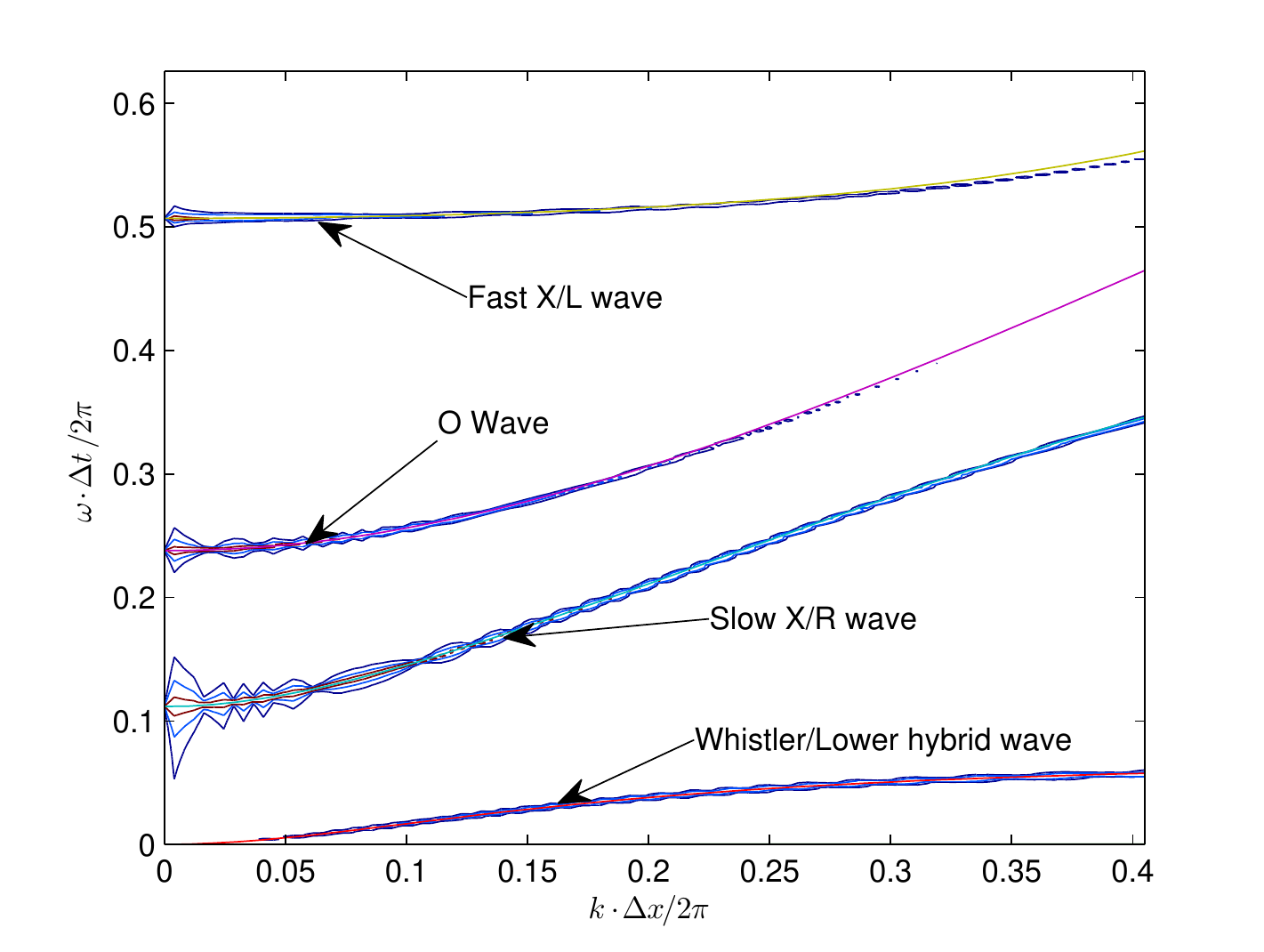}

}

\subfloat[Low frequency region.]{\includegraphics[width=0.49\textwidth]{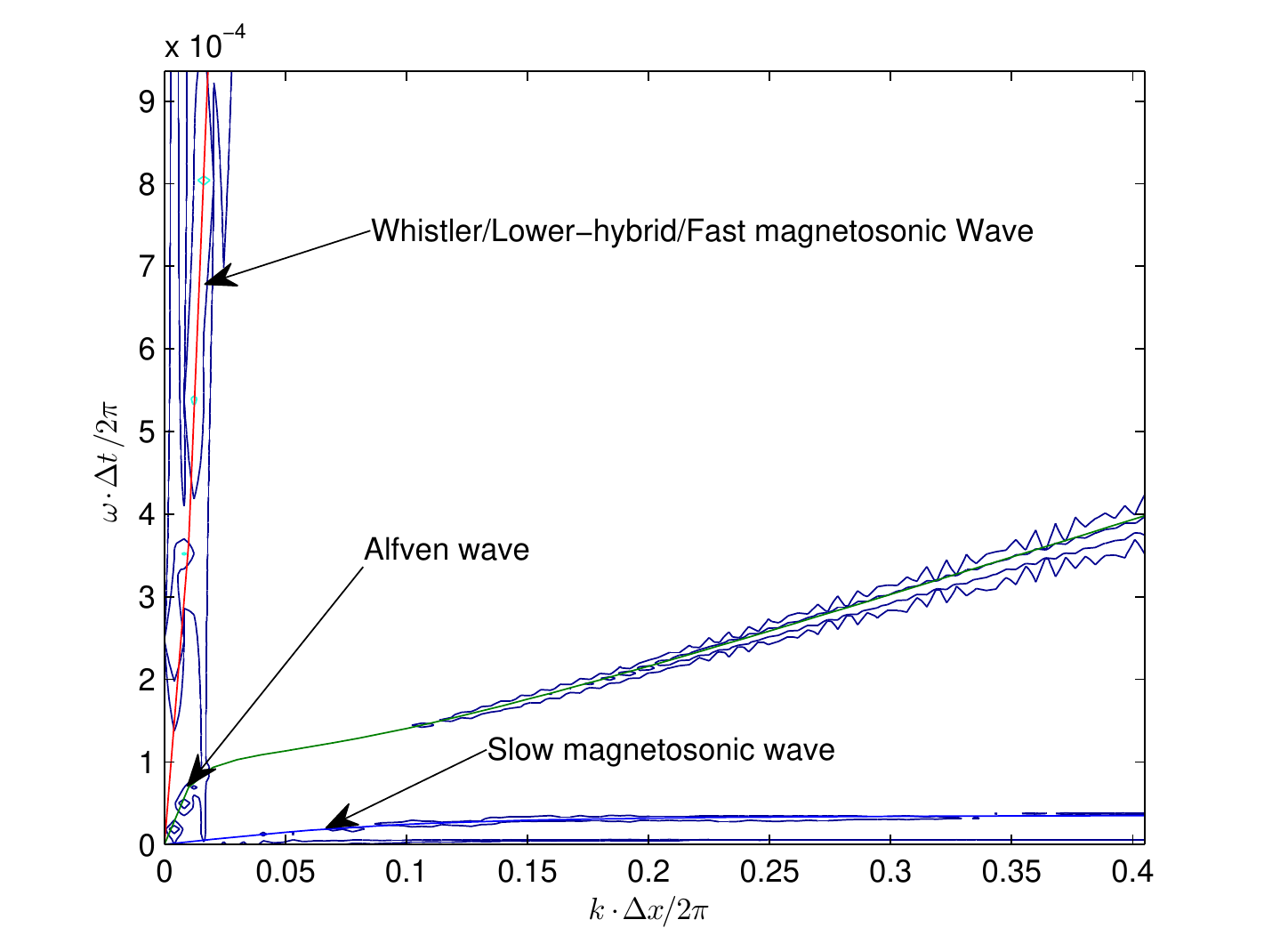}

}

\caption{The dispersion relation of an electron-deuterium two-fluid plasma
plotted against simulation results. The high frequency region is shown
in (a) and the low frequency region in (b), with solid lines being
theoretical dispersion relation. }

\label{FigDISPREL} 
\end{figure}

\begin{figure}
\includegraphics[width=0.6\textwidth]{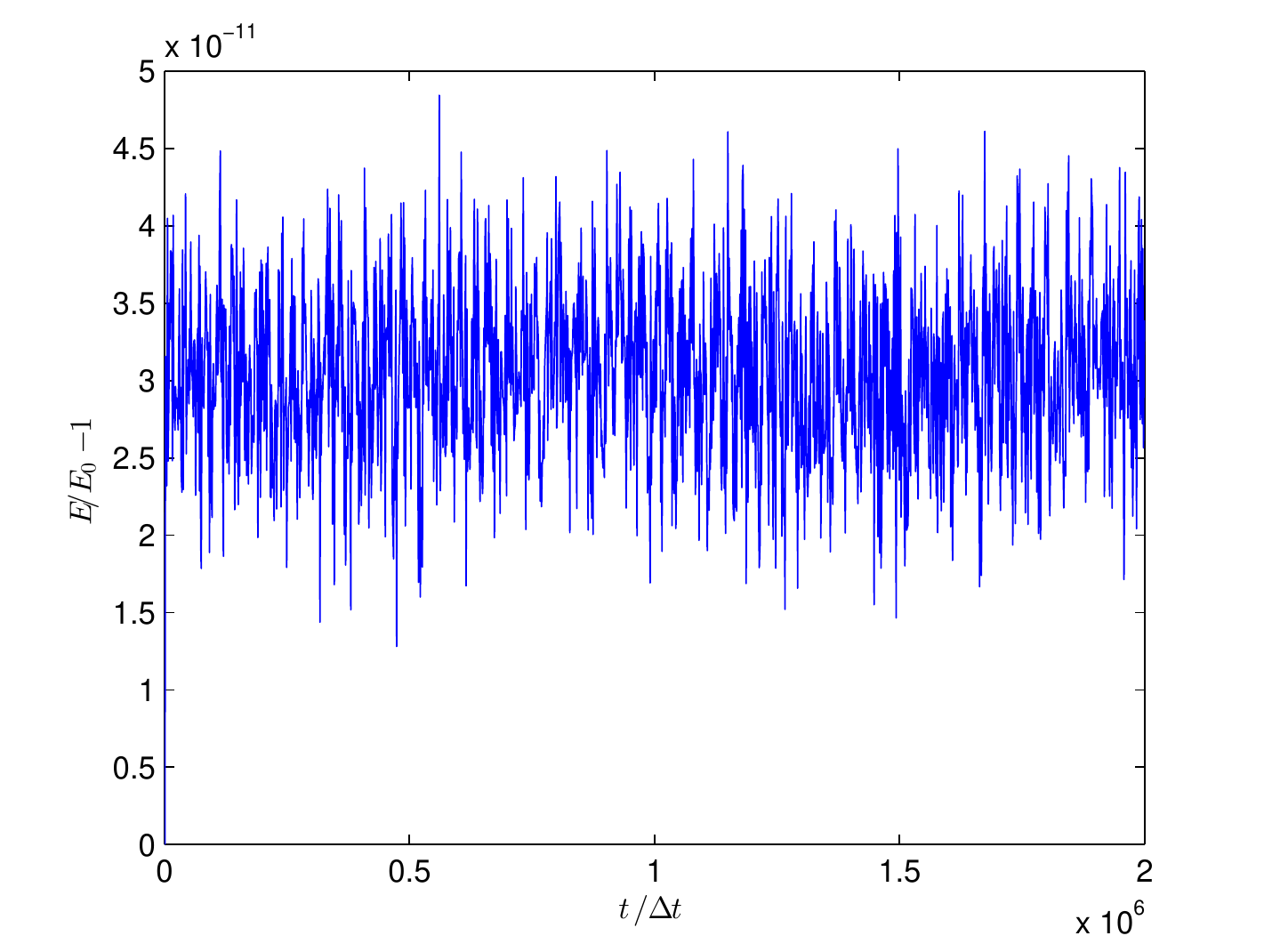}

\caption{Evolution of the total energy obtained from the structure preserving
two-fluid algorithm. In all $2\times10^{6}$ time steps the energy
deviates from its initial value by only a very small amount.}

\label{FigENE} 
\end{figure}

To numerically obtain the dispersion relation, the simulation is carried
out with a small random perturbation. The space-time dependence of
one field component is transformed into $\omega-k$ space, and a contour
plot the field component in the $\omega-k$ space is used to make
correspondence with the linear dispersion relation of the discrete
system. In Fig.~\ref{FigDISPREL}, such a contour plot is compared
with the theoretical dispersion relation in both high frequency and
low frequency ranges. We can see that the dispersion relation obtained
by our geometric two-fluid algorithm agrees very well with the theory
over the frequency range of the simulation. As expected, the total
energy of the system is bounded to be within a interval of its intitial
value of for all simulation time-steps, which is plotted in Fig.~\ref{FigENE}.

\begin{figure}
\subfloat[Stable mode for $k_{z}=\pi/32$.]{\includegraphics[width=0.49\textwidth]{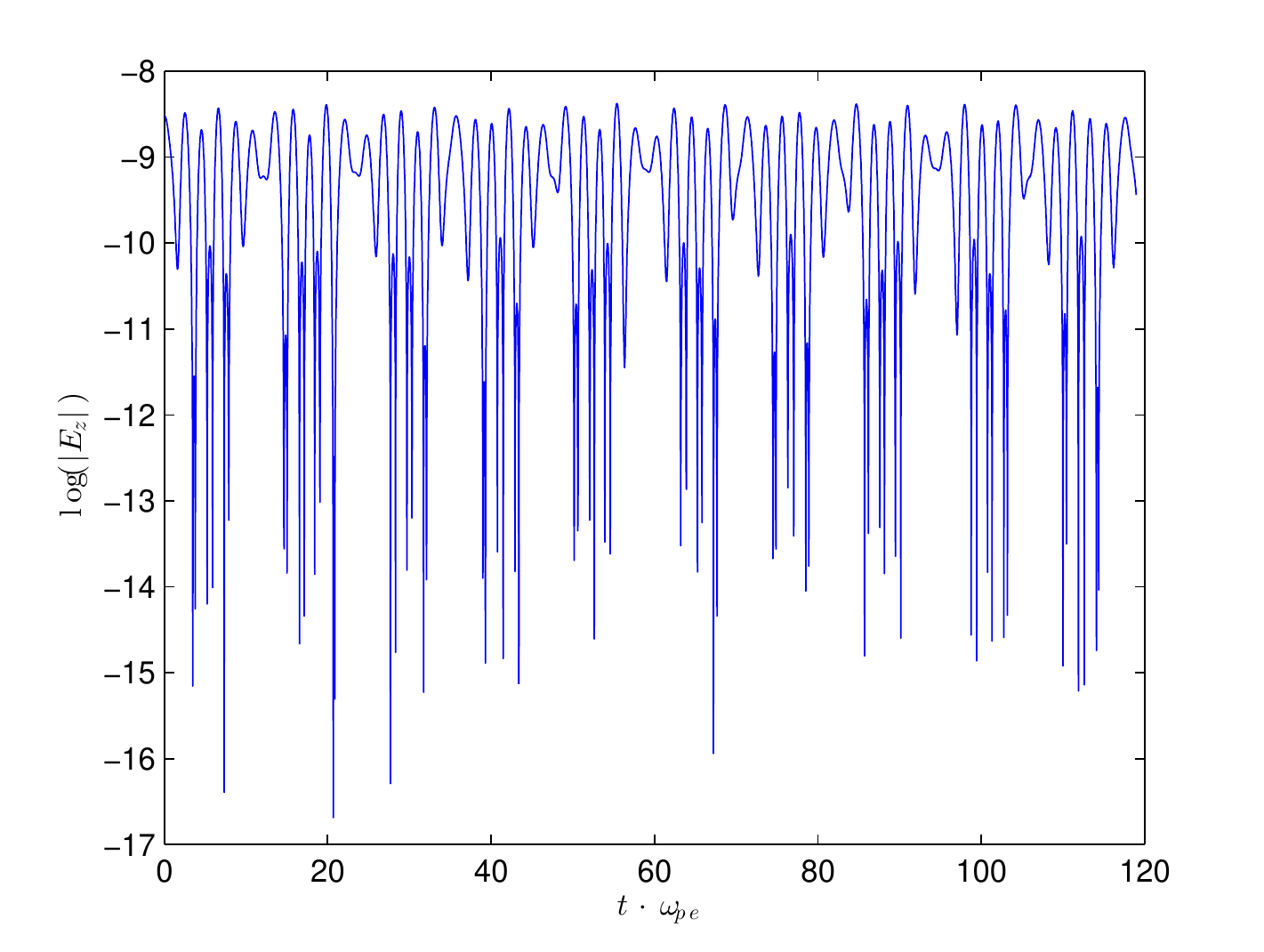}

}

\subfloat[Unstable mode for $k_{z}=\pi/128$.]{\includegraphics[width=0.49\textwidth]{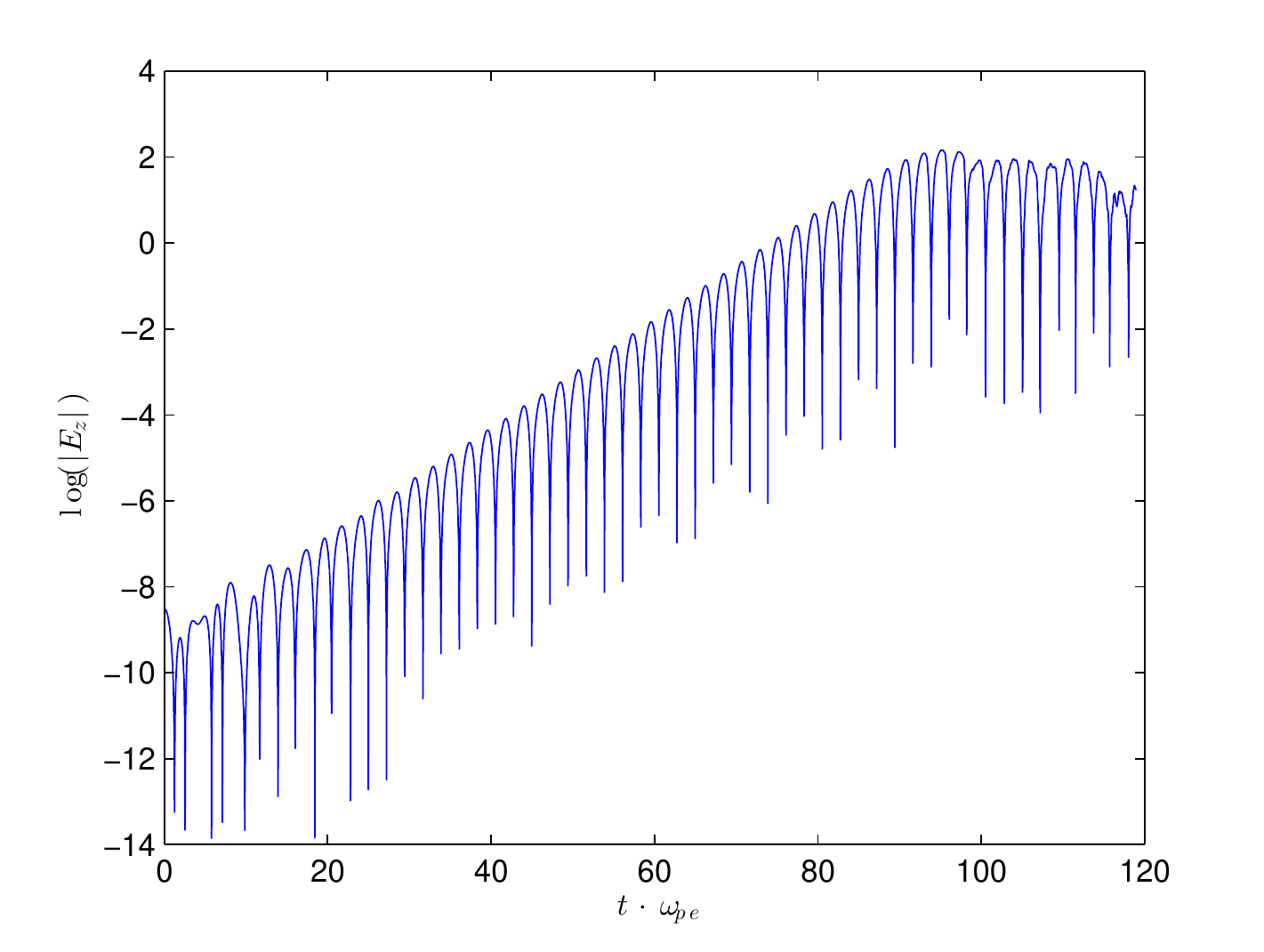}

}

\caption{Simulation of the oscillating two-stream instability for an electron-positron
plasma. Simulations show that the mode with $k_{z}=\pi/32$ is stable
(a) while the $k_{z}=\pi/128$ is unstable (b), as predicted theoretically
in Ref.~\cite{qin2014two}.}
\label{FigTSinstability} 
\end{figure}

\begin{figure}
\includegraphics[width=0.6\textwidth]{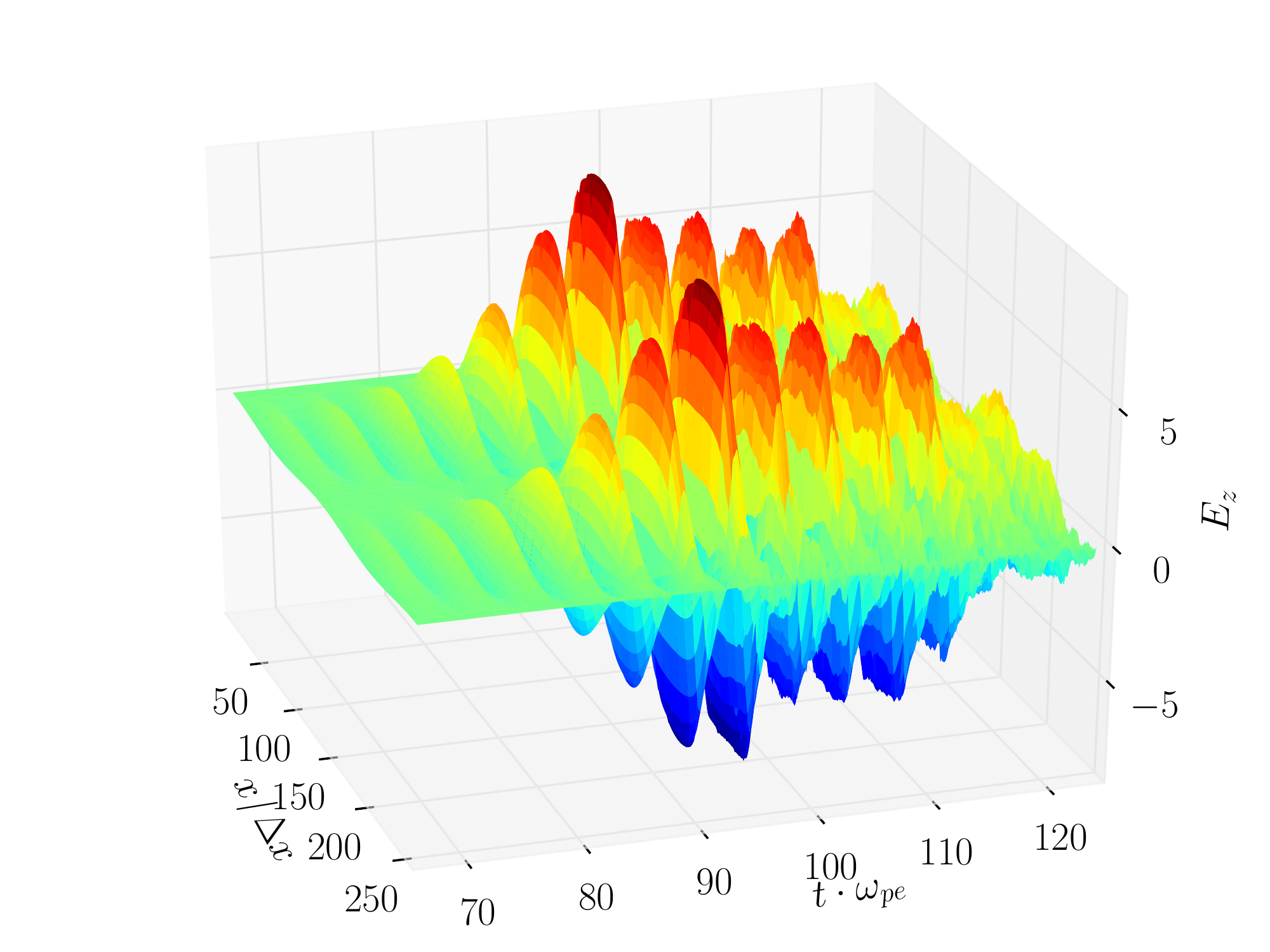}

\caption{The space-time dependence of $E_{z}$ during the nonlinear evolution
of the unstable mode initiated by a perturbation with $k_{z}=\pi/128$
.}

\label{FIGTSEz} 
\end{figure}

The second example is the well-known oscillating two-stream instability
\cite{nishikawa1968parametric,morales1974nonlinear}. We consider
the case of an unmagnetized cold two-fluid model, and compare with
stability condition that was previously studied in Ref.~\cite{qin2014two}.
We simulate an electron-positron plasma, with system parameters given
as follows: 
\begin{eqnarray}
n_{e0}=n_{i0} & = & 4.0\EXP^{15}\mathrm{m}^{-3}~,\\
m_{i}=m_{e} & = & 9.1\EXP^{-31}\mathrm{kg}~,\\
\bfB_{0} & = & 0~,
\end{eqnarray}
wiht the relative drift velocity between electrons and positrons chosen
to be $\bfv_{d}/2=\bfv_{e0}=-\bfv_{i0}=\bfe_{z}0.041\mathrm{c}$.
The simulation domain is a $1\times1\times256$ mesh. Initial perturbations
with two different wave numbers, $k_{z}=\pi/128$ and $k_{z}=\pi/32$,
are tested. According to the theoretical prediction of Ref.~\cite{qin2014two},
the mode with $k_{z}=\pi/32$ is stable while that with $k_{z}=\pi/128$
is unstable. Both of these predictions are confirmed by the simulation
using our algorithm, as seen in \FIG{FigTSinstability}. The evolution
of the perturbed electrostatic field $E_{z}$ of the unstable mode
is plotted in \FIG{FIGTSEz}, which displays the space-time dependence
of $E_{z}$ during the nonlinear evolution of the instability.

\section{Conclusion}

\label{SecDis}

In this paper, a geometric structure preserving algorithm for ideal
two-fluid systems was developed. In this method, fluids were discretized
as Lagrangian particles, and the conservation of mass was seen to
be naturally satisfied. The electromagnetic and internal energy fields
were discretized over a fixed cubic mesh using discrete differential
forms. With the help of high-order Whitney interpolation forms, this
scheme preserves the electromagnetic gauge symmetry. In the algorithm
the discrete pressure was obtained from the discrete internal energy
field. The time integration was accomplished by adopting a powerful
high-order explicit Hamiltonian splitting technique, which preserves
the whole symplectic structure of the two-fluid system. Numerical
examples were given to verify the accuracy and conservative nature
of the geometric algorithm. We expect this algorithm will find a wide
range of applications, especially in physical problems that are multi-scale
and demand long-term accuracy and fidelity.
\begin{acknowledgments}
This research is supported by ITER-China Program (2015GB111003, 2014GB124005,
2013GB111000), JSPS-NRF-NSFC A3 Foresight Program in the field of
Plasma Physics (NSFC-11261140328), the National Science Foundation
of China (11575186, 11575185, 11505185, 11505186), the CAS Program
for Interdisciplinary Collaboration Team, the Geo-Algorithmic Plasma
Simulator (GAPS) project. PJM was supported by U.~S.~Dept.~of Energy
contract No.~DE-FG02-04ER-54742. 
\end{acknowledgments}

 \bibliographystyle{apsrev4-1}
\bibliography{geotfnew}

\end{document}